\documentclass[aps,prl,showpacs,preprintnumbers,linenumbers,english,twocolumn,amsmath,amssymb]{revtex4}
\usepackage[T1]{fontenc}
\usepackage[latin1]{inputenc}
\usepackage{graphicx}
\usepackage{amssymb}
\usepackage{verbatim}
\usepackage{babel}
\usepackage{lmodern}
\usepackage{color,soul}
\usepackage{epsfig}
\usepackage{txfonts}
\usepackage{subfig}
\usepackage{float}
\usepackage{wasysym}
\usepackage{url}
\usepackage{hyperref}
\usepackage{rotating}
\hypersetup{breaklinks=true}
\newcommand{\pu}{\ensuremath{^{239}}Pu\,}
\newcommand{\cf}{\ensuremath{^{252}}Cf\,}
\newcommand{\ein}{\ensuremath{E_{in}^{n}}\,}

\newcommand{\nubar}{\ensuremath{\overline{\nu}_{p}\,}}

\newcommand{\jeff}{JEFF3.3 }
\newcommand{\ENDF}{ENDF/B-VIII.0 }

\newcommand{\citescintitanks}{ \cite{conde,diven1962,frehaut,gwinUpTo10MeV,Khokhlov,Mather1965,savin,Smirenkin1958,soleilhac,Walsh1971}}
\newcommand{\citeparaffineblock}{\cite{Flerov1961,Hopkins1963,jacob,johnstone56,Kalashnikova,leroy1960,nesterov,Nurpeisov1975,Volodin}}

\date{}
\begin{document}

\title{Energy Dependence of Prompt Fission Neutron Multiplicity in the $^{239}$Pu($n,f$) Reaction}

\author{P.~Marini$^{1,2}$}\email{marini@cenbg.in2p3.fr}
\author{J.~Taieb$^{1,3}$}
\author{D.~Neudecker$^{4}$}
\author{G.~B\'elier$^{1,3}$}
\author{A.~Chatillon$^{1,3}$}
\author{D.~Etasse$^{5}$}
\author{B.~Laurent$^{1,3}$}
\author{P.~Morfouace$^{1,3}$}
\author{B.~Morillon$^{1}$}
\author{M.~Devlin$^{4}$}
\author{J.~A.~Gomez$^{4}$}
\author{R.~C.~Haight$^{4}$}
\author{K.~J.~Kelly$^{4}$}
\author{J.~M.~O'Donnell$^{4}$}

\affiliation{
$^{1}$CEA, DAM, DIF, F-91297 Arpajon, France\\
$^{2}$Univ. Bordeaux, CNRS, CENBG, UMR 5797, F-33170 Gradignan, France\\
$^{3}$Universit\'e Paris-Saclay, CEA, LMCE, 91680 Bruy\`eres-le-Ch\^atel, France \\
$^{4}$
Los Alamos National Laboratory, Los Alamos, NM-87545, USA\\
$^{5}$Normandie Univ, ENSICAEN, UNICAEN, CNRS/IN2P3, LPC Caen, 14000 Caen, France 
}

\date{\today}

\begin{abstract}
Accurate multiplicities of prompt fission neutrons emitted in neutron-induced fission on a large energy range are essential for fundamental and applied nuclear physics. 
Measuring them to high precision for radioactive fissioning nuclides
remains, however, an experimental challenge. 
In this work,  the average prompt-neutron multiplicity emitted in the \pu($n,f$) reaction was extracted as a function of 
the incident-neutron energy, over the range 1-700~MeV, with a novel technique, which  
allowed 
to minimize and correct for the main 
sources of bias and thus achieve unprecedented precision.

At low energies, our data validate for the first time the \ENDF nuclear data evaluation with an independent measurement and reduce the evaluated uncertainty by up to $60\%$.
 This work opens up the possibility of precisely measuring prompt fission neutron multiplicities on highly radioactive nuclei relevant for an essential component of energy production world-wide.\\

\end{abstract}

\maketitle

Despite the discovery of nuclear fission being 80 years old, a full understanding of this rich quantum phenomenon  is still a challenge for experimentalists and theoreticians. Parallel efforts \cite{Schmidt_2018,
endf8,JEFF33}, pursued worldwide,  carry the promise of a renewed understanding of this complex phenomenon, and support the development of modern nuclear technologies for energy production effectively complying with the most recent requirements on safety, sustainability, economic competitiveness and proliferation resistance. 
From an experimental point of view, the most stringent constraints to theoretical models  are expected to come from very precise measurements of observables over large energy ranges,
as well as from the simultaneous measurement of several observables highlighting their possible correlations.
Among them, the number of prompt fission neutrons and their kinetic energy distributions 
provide 
valuable information on the amount of excitation energy of the heated fissioning system 
transferred to the primary fragments.
Moreover, these data, for the fissile $^{235}$U and \pu isotopes and the fertile $^{238}$U nuclide, are vital inputs to calculate next-generation nuclear reactor neutronics, which affect projections of the criticality, efficiency, safety, and lifetime of such systems. 
From a theoretical point of view, a model able to describe the fission process with the requested accuracy is still lacking, therefore nuclear data applications  rely, to a large extent, on evaluated data, such as  \ENDF and \jeff \cite{endf8, JEFF33}.

Several experimental and evaluation works have been dedicated since the '60-'70s to produce coherent and precise data and evaluations for the average prompt fission neutron multiplicity  (\nubar) emitted in the neutron-induced fission of \pu in the MeV range 
\cite{bethe1955,Kalashnikova,Kolosov,nesterov,Nurpeisov1975,Volodin,
Smirenkin1958,leroy1960,Flerov1961,Hopkins1963,johnstone56,Walsh1971,Mather1965,conde,frehaut,savin,soleilhac,gwinUpTo10MeV,diven1962,jacob,Khokhlov,wang2019}. 
 The importance of these data lies in the fact that the sustainability of the nuclear fission chain reaction in a reactor core, the so-called criticality $k_\mathrm{eff}$, depends nearly linearly on the 
\nubar 
  of the fissioning nuclide \cite{reuss} and 
often has the highest sensitivity to the \nubar  of the main fuel \cite{dice}.
 The precise measurement of \nubar for a highly radioactive nuclide, as  \pu, is, however, an experimental challenge as it requires an unambiguous identification of fission events from a very intense $\alpha$-decay background.
Moreover, neutrons need to be detected with good efficiency and discriminated from $\gamma$ rays emitted in fission. 
The current widely accepted reference measurement of J. Fr\'ehaut et al. \cite{frehaut}, carried out in the 70's,  provides the 
data reported to be most precise in 
 the incident-neutron energy region between 1 and 30 MeV, with reported uncertainties as low as $0.5\%$   below $15\,$MeV.
 Other experimental data \cite{bethe1955,Kalashnikova,Kolosov,nesterov,Nurpeisov1975,Volodin,Smirenkin1958,leroy1960,Flerov1961,Hopkins1963,johnstone56,Walsh1971,Mather1965,conde,frehaut,savin,soleilhac,gwinUpTo10MeV,diven1962,jacob, Khokhlov,wang2019},
 although with larger uncertainties, are all in good agreement with this measurement.
 The large majority and the most precise of these measurements 
  \citescintitanks \, were realized detecting neutrons in coincidence with fission events with a close-to-$4\pi$ scintillator detector tank.
%
Other measurements exploited either proportional counters inside paraffin blocks
\citeparaffineblock \, or the surrogate-reaction technique \cite{wang2019}.

\begin{figure}[t]
\centering
\includegraphics[width=1\columnwidth,clip]
{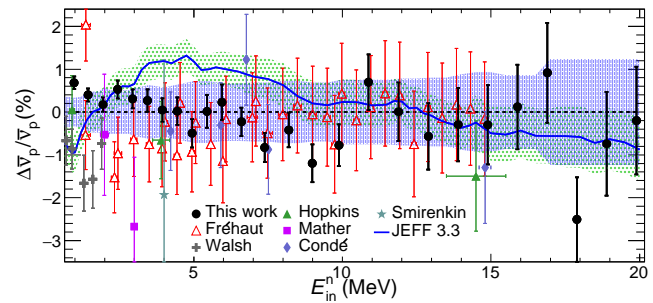}
\caption{(Color online) Relative difference between experimental and \ENDF \nubar values
from this work and 
some previous experiments \cite{
Smirenkin1958,
Hopkins1963, Mather1965,conde,
frehaut,Walsh1971
}. 
The difference with
 the \jeff evaluation \cite{JEFF33} is also shown.} 
\label{fig:mnEtDonnees} 
\end{figure}

Resulting libraries are then validated with respect to integral experiments such as, e.g., $k_\mathrm{eff}$ experiments   \cite{icsbep} that model the behavior of reactor cores on a small scale. This validation steps tests the reliability of entire libraries for applications.
\ENDF \pu(n,f) \nubar  were obtained from existing experimental data, but the evaluated data had to be adjusted \cite{endf8} such that simulated and experimental $k_\mathrm{eff}$ were in reasonable agreement for application calculations.
%
To illustrate this point, in Fig.\ref{fig:mnEtDonnees} we show the relative difference between existing experimental and \ENDF \nubar values, with the data normalized to the current \ENDF value of \cf \nubar \cite{DenisePrivateComm} and not including its uncertainty. 
For a more readable figure, the uncertainty on the \ENDF values was not propagated in the relative difference and 
is shown as shaded region around the zero value.
 Discrepancies 
as high as $2\%$ below 8 MeV are observed, with data systematically lower than the most recent evaluation.  
A different trend and differences up to $1\%$ are also observed for the \jeff evaluation.
As a comparison, it should be kept in mind that a change in \nubar by 0.1\% in an energy range as small as $100\,$keV can modify the computed criticality by about 100 pcm, which is about one third of the range between a controlled 
and an uncontrolled  Pu critical assembly \cite{Neudecker_2020,Neudecker_2021}.

In this letter we report on experimental data, obtained  to high precision with a different and novel technique.  
They validate, below 5 MeV, the \ENDF evaluation with an independent measurement, highlight potential shortcomings in existing data, and reduce the evaluated uncertainty of up to $60\%$ while extending the range of the studied energies from $30$ to $700\,$MeV.
The method, used here for \pu($n,f$) \nubar measurement for the first time, consists of detecting neutrons with an ensemble of 17.78 cm-diameter scintillator detectors, located on a half-sphere at about 1 meter distance from a fission detector assuring a good energy resolution for the emitted neutrons. Prompt fission neutron spectra (PFNS) are then measured,  as a function of the incident-neutron energy, with a double time-of-flight technique  \cite{Ethvignot_PRL94nubarU8U5}.
Values of \nubar are finally extracted from the integration of PFNS. 
As opposed to 
tank experiments \citescintitanks, these kinds of measurements suffer from  statistics limitations, due to the limited detector angular coverage, 
and from the presence of a background due to neutron scattering on surrounding materials. 
However, they present two main advantages. First, while scintillator tanks are, at most, roughly segmented,
the high segmentation of the neutron detector array allows to measure the neutron angular distribution, and therefore to precisely correct for the contribution of regions not covered by the detector.
Second, PFNS are precisely measured and  an energy-dependent efficiency curve, typically determined with respect to a \cf source, can be used. This is not the case in  tank experiments, where only the difference in the mean energy of the $^{252}$Cf  and \pu PFNS could be accounted for, based on an empirical parametrization \cite{terrel1962}, insufficient to today's standards. These two effects lead to systematic biases in the existing \nubar measurements not properly corrected for, thus enlarging further the level of uncertainty on \nubar.
On the contrary, in our experiment
the availability of a high-intensity, pulsed and well-collimated white neutron source, and the described novel technique 
  allowed for the first time to effectively minimize and estimate 
  the sources of possible bias 
  while collecting high counting statistics  and providing an independent measurement.

The experiment was performed at the Weapons Nuclear Research facility \cite{wnr1,wnr} of Los Alamos Neutron Science Center at the Los Alamos National Laboratory. 
The neutron beam 
was produced by spallation and 
bombarded a high-purity \pu target after a flight path 
of about $21.5\,$m. 
A newly-developed, high-efficiency, light-weight, fast fission chamber, with an improved 
discrimination capability between fission and $\alpha$-decay events \cite{laurent2021}, was coupled to 54 EJ-309 \cite{eljen_EJ} liquid scintillators from the Chi-Nu array \cite{chinu} to detect neutrons emitted in fission events.
The fission chamber housed 47 mg of \pu arranged in twenty-two deposits and eleven readout channels, with an $\alpha$-activity of about $10\,$MBq per channel, to be compared to a fission rate of about $15$ events/s.
A fission-fragment detection efficiency of $95\%$ was nevertheless achieved \cite{laurent2021}. Such a feature is crucial
to avoid any bias of the data associated with the selection of a particular range
in angle or kinetic energy of the detected fragments.
Neutrons and $\gamma$ rays were detected in coincidence with a fission-chamber signal in the  scintillator cells 
and identified via the pulse shape discrimination technique down to $200\,$keV and up to about $14\,$MeV. The neutron detectors covered nine angles, from $30^{\text{o}}$ to $150^{\text{o}}$. 
The use of digital Fast Acquisition SysTem for nuclEar Research \cite{faster} allowed the near complete avoidance of numerical dead time.
A detailed description of the experimental setup can be found in \cite{mariniPFNS_PuWNR, taieb, laurent2021, chinu}.

 The combined setup and the high recorded statistics  lead to a  precise reconstruction of the PFNS 
 as a function of the incident-neutron energy \ein, from 0.7 to 700$\,$MeV. The analysis of the experimental data and the associated uncertainties are discussed in \cite{mariniPFNS_PuWNR}. 
Here we only recall that 
neutron detector efficiencies were obtained by measuring the PFNS of the \cf spontaneous fission reaction in the same experimental conditions and with the same analysis procedure as the \pu. 
For each detector, the measured \cf($sf$) PFNS was divided by the 
evaluated PFNS standard \cite{carlson2009}, normalized to the evaluated \ENDF \nubar of ($3.759\pm0.42\%$) \cite{endf8}, and used to evaluate the efficiency of every EJ-309 detector. 
 The bias associated with this procedure was carefully evaluated via GEANT4 simulations \cite{geant4} and found to be negligible \cite{mariniPFNS_PuWNR}.
The prompt fission neutron spectrum for each of the eighty-six \ein bins studied 
was obtained by combining all the detector spectra corrected for their efficiency.
As each of them was corrected by its neutron detector efficiency, the integral of the  PFNS is the average number of prompt neutrons emitted per fission (\nubar).  
Results are presented with the absolute statistical and systematic uncertainties, propagated through the data analysis. 
The latter includes the uncertainty on the evaluated $^{252}$Cf PFNS, while the uncertainty of $0.42\%$ on the \cf \nubar is not included. This will allow to easier account for more precise future measurements of \cf \nubar.

\begin{figure}[t!]
\centering
\includegraphics[width=1\columnwidth,clip]{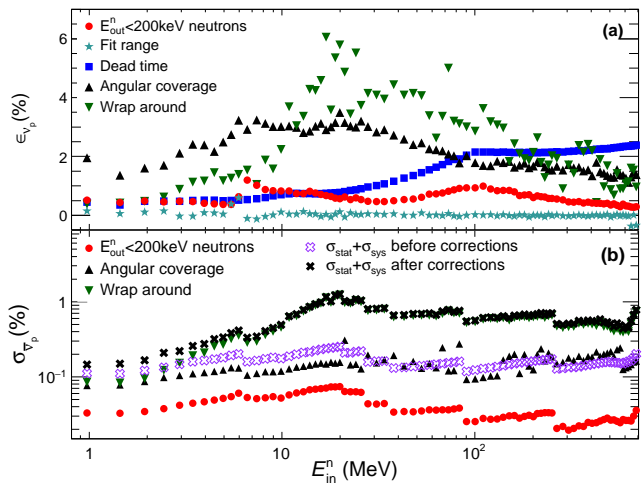}
\caption{(Color online) (a) Bias of \nubar due to different experimental effects. (b) Uncertainty on \nubar ($\sigma_{\overline{\nu}_{p}}$) introduced by the correction of the experimental bias and the sum of the $\sigma_{\overline{\nu}_{p}}$ before and after the corrections. The ``fit range'' and the ``dead time'' corrections $\sigma_{\overline{\nu}_{p}}$ are negligible (see text).}
\label{fig:corrections} 
\end{figure}

The main breakthrough with respect to previous measurements is the possibility of effectively  estimating 
  the sources of possible systematic bias. 
Data were corrected for four different experimental biases:  the neutron detection energy range, the presence of a slower incident neutron background (wrap-around)  \cite{wraparound}, 
the limited detector angular coverage and the detector dead time. 
The correction on the \nubar values related to each of these physical effects ($\epsilon_{\overline{\nu}_{p}}$),
 as well as the uncertainty introduced on the final \nubar value
by each correction ($\sigma_{\overline{\nu}_{p}}$), 
%
are plotted in Figs. \ref{fig:corrections}a and b, respectively, as a function of \ein. 

First, the detection limits of $0.2$ and $14\,$MeV 
for the fission neutrons were considered. 
The lower limit was set by the threshold for 
discriminating neutrons from $\gamma$-rays, 
while the 
high-energy one was related to the dynamic range of the used electronics. 
For corrections needed by these detection limits, the two regions were handled separately. 
The contribution to \nubar of neutrons below the $200\,$keV detection limit  was estimated assuming a simple theoretical description of the low-energy region of the PFNS based on a Maxwellian spectrum \cite{bloch1943,terrell1959} and found to be as high as $1\%$, as shown in Fig. \ref{fig:corrections}a (red dots). The measured spectra were therefore extrapolated at energies from $200\,$keV downwards 
and \nubar corrected for it. The bias introduced by the arbitrary choice of the fitting range 
on the extracted \nubar 
found to be smaller than $0.2\%$ (cyan stars in Fig. \ref{fig:corrections}a). The uncertainty introduced on the \nubar values by this procedure is negligible with respect to the statistical and systematic uncertainty of the  uncorrected value (compare red dots and open crosses in Fig. \ref{fig:corrections}b).
For the high-energy limit, we can reasonably expect neutrons above $14\,$MeV to be emitted during a pre-equilibrium, pre-fission process for incident energies above about $24\,$MeV. A TALYS calculation \cite{talys} estimates their contribution to be about $0.15\%$ of \nubar at this \ein.  
It should be noted that this process can contribute significantly to \nubar, but it cannot be estimated quantitatively as available pre-equilibrium emission models have been validated on limited experimental data \cite{kawano}. Therefore, \nubar values for \ein above $24\,$MeV should be considered as a lower limit.

Second, the obtained values of \nubar were corrected for neutrons emitted in fissions 
induced by slower-than-measured neutrons, the wrap-around background \cite{wraparound}. 
The fraction of wrap-around background, $k_{WA}$, in each \ein  bin could be analytically determined from the time-of-flight spectra of incident neutrons as described in \cite{mariniPFNS_PuWNR}. The procedure
was validated by the observation of dips in the evolution of $k_{WA}$ with \ein at energies corresponding to absorption resonances in $^{16}$O, $^{14}$N (i.e. air) and $^{11}$B (boron material present in the beam hardener).
The $k_{WA}$ fraction varies from about $10\%$ to about $3\%$ below 20 MeV and above 200 MeV, respectively, of the impinging neutron flux and modifies the \nubar value up to $6\%$, pointing out its importance (green triangles in Fig. \ref{fig:corrections}a). 
The relative uncertainty introduced on the \nubar values by the correction of this effect, which reaches up to $1\%$, arises from the statistics available for the estimation of $k_{WA}$ (green triangles in Fig. \ref{fig:corrections}b).

\begin{figure}[t!]
\centering
\includegraphics[width=1\columnwidth,clip]
{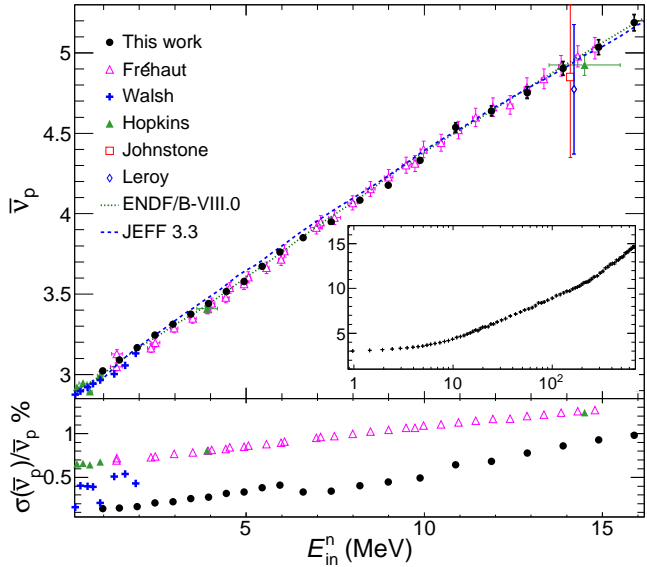}
\caption{(Color online) Measured \nubar and its uncertainty as a function of incident neutron energy up to $16\,$MeV. Some data from previous experiments are also shown \cite{frehaut,Walsh1971,Hopkins1963,leroy1960,johnstone56}. The \cf \nubar uncertainty was removed from existing data. Dotted and dashed lines are \ENDF and \jeff evaluations, respectively. 
The insert shows the measured \nubar over the whole studied \ein energy range.
  }
\label{fig:mn 8MeV} 
\end{figure}

Third, the limited detector angular coverage was considered.
The high segmentation of the Chi-Nu array allowed for the reconstruction of the \nubar angular distribution and the correction for those angles that were not covered by detectors.
Nine spectra, one for each measured $\theta_{lab}$, were obtained by combining the spectra 
from the six detectors at the considered angle and \nubar$(\theta,E_{in})$ extracted.
Their uncertainty is close to $0.3\,$\%. 
The angular distributions, \nubar$(\theta,E_{in})$ vs cos$(\theta)$,
exhibit two main characteristics: first, they are not isotropic, even at low incident energies, with a \nubar$(0^{\circ})/$\nubar$(90^{\circ})$ of about $1.05$ below $10\,$MeV, and a trend similar to the one observed in the data for fission-fragment anisotropy \cite{Simmons1960,shpak,vorobyev2018}. 
 Second, they  are characterized by a certain degree of forward/backward asymmetry which increases with \ein, 
 reflecting the increase in the  kinematical boost and  the pre-equilibrium emission.
Angular distributions were  fitted with up to 4$^{th}$-order polynomial functions and  \nubar$(E_{in})$ 
reconstructed as described in \cite{mariniPFNS_PuWNR} for the PFNS mean energy.
The systematic uncertainty on \nubar \, due to the arbitrary choice of the functions was found to be negligible with respect to its final uncertainty.
Accounting for the neutron angular distribution modifies  up to $4\%$ the \nubar values and it mainly arises from the contribution of the most forward/backward angles 
(black triangles in Fig. \ref{fig:corrections}a). This implies that 
the assumption of a flat angular distribution or a non-accurate knowledge of it likely leads, even at low energies, to an underestimation of \nubar. 
Interestingly, existing literature data are generally lower than the present results, and this could 
be a source of discrepancy (see Figs.\ref{fig:mnEtDonnees} and \ref{fig:mn 8MeV}).
The uncertainty on the \nubar values introduced by the described correction 
is shown as black triangles in Fig.\ref{fig:corrections}b.

Finally, the impact of the neutron
detector dead time was investigated. 
Once a particle (neutron or $\gamma$ ray) fires a scintillator detector of the Chi-Nu array, a charge-integration window of $200\,$ns is opened, during which any other impinging particle is not recorded separately, but its charge signal adds to that of the first one. Therefore particles impinging with a time difference smaller than $200\,$ns in the same detector can be mis-identified and neutrons can be ``lost'' or ``gained''.
The net amount of ``lost'' neutrons was estimated with a full Monte Carlo simulation based on experimental distributions. Measured neutron and $\gamma$-ray multiplicities, as well as time-of-flight distributions for each \ein bin were sampled and used as input. A similar procedure was undertaken for the fast-to-total signal charge ratio vs total signal charge correlation 
of each detector \cite{mariniPFNS_PuWNR}. 
The same simulation, with the appropriate inputs, was run for the $^{252}$Cf($sf$) data, as part of the detector dead time distortion is accounted for when correcting the PFNS for the detector efficiency. The net correction varies from $\sim0.5\%$ to above $2\%$ for energies below $10\,$MeV and above $100\,$MeV, respectively (squares in Fig. \ref{fig:corrections}a), due to the increase of $\gamma$ and neutron multiplicities as the incident-neutron energy increases. 
The number of simulated events is high enough so that  the statistical uncertainty introduced by this correction is negligible.

The data corrected as described above are shown 
as a function of \ein in Figs. \ref{fig:mnEtDonnees} and
\ref{fig:mn 8MeV}. 
Our data exhibit the expected constant increase up to $700\,$MeV with no fluctuations.
As mentioned, \nubar values for energies above about 24 MeV, where the contribution of high-energy ($>14$MeV) pre-equilibrium neutrons becomes non negligible, should be considered as a lower limit. 
Below about $14\,$MeV, \nubar exhibits a linear dependence with the neutron energy.
A linear extrapolation below 3 MeV provides an estimated value of \nubar at thermal neutron energies
of  ($2.879 \pm 0.010$) neutrons/fission, in 
 agreement with the evaluated value of ($2.868\pm0.012$) \cite{endf8} and with comparable uncertainty.

The obtained \nubar total uncertainties (without $0.42\%$ uncertainty of \cf \nubar) span from $0.15$ to $1.3\%$, and are smaller than $1\%$ below 14 MeV \ein (see Figs. \ref{fig:corrections}b and \ref{fig:mn 8MeV}). 
Such low uncertainties on a broad energy range were never reached before, not even with different experimental techniques (
  \cite{conde,diven1962,frehaut,gwinUpTo10MeV,jacob,Khokhlov,Mather1965,savin,soleilhac,Walsh1971} 
  and
  \cite{Hopkins1963,Kalashnikova,Volodin}) as shown in the bottom panel of Fig. \ref{fig:mn 8MeV}. 
The relative difference between our data and \ENDF values is shown in Fig. \ref{fig:mnEtDonnees}. 
At low \ein our data show a different trend than the reference experimental data of J.~Fr\'ehaut et al. \cite{frehaut}, but 
agree, on average better than $0.3\%$ below 8 MeV, with the recent \ENDF evaluation \cite{endf8}. 
A significant discrepancy is observed at the opening of the second-chance fission.

\begin{figure}[t!]
\centering
\includegraphics[width=1\columnwidth,clip]
{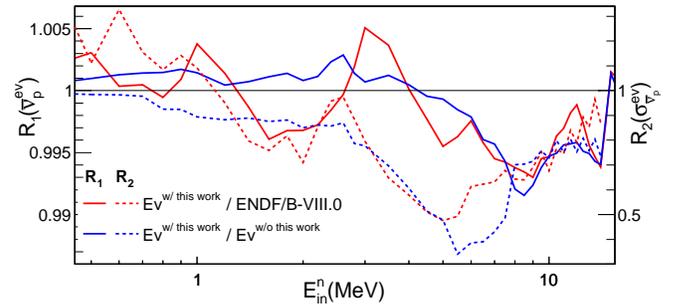}
\caption{(Color online)
Ratio of \nubar evaluated mean value (R$_{1}$, full lines, left axis) and relative uncertainty (R$_{2}$, dashed lines, right axis) for different evaluations (see text).
}
\label{fig:evaluation} 
\end{figure}

To asses the impact of our data compared to existing data sets, we performed two new evaluations of \pu \nubar, with (\textit{Ev$^{w/\,this\,work}$}) and without the data presented here (\textit{Ev$^{w/o\,this\,work}$}), using the same methodology.
The  uncertainties of the all data 
 \cite{Nurpeisov1975,Volodin,Hopkins1963,Walsh1971,Mather1965,conde,frehaut,savin,soleilhac,gwinUpTo10MeV,diven1962,Khokhlov}  
 were  carefully reviewed and increased according to \cite{Neudecker_TemplateUnc} in cases where 
 uncertainties 
 were missing. 
Our data reduce the \ENDF and \textit{Ev$^{w/o\,this\,work}$} \nubar evaluated relative uncertainty, \ensuremath{\sigma_{\overline{\nu}_{p}}^{ev}\,},  by up to $50\%$ and $60\%$, respectively, in the 1 to 15 MeV range (see dashed lines in Fig.\ref{fig:evaluation}). This is of high importance for nuclear applications as it reduces the uncertainties and increases the predictive power of neutronics calculations.
 %
In addition to the high-impact due to low uncertainties, our data offer two additional benefits.
Below $5\,$MeV the \nubar evaluated mean value, \ensuremath{\overline{\nu}_{p}^{ev}\,}, is only sligthly modified ($<0.15\%$) by our data (Fig.  \ref{fig:evaluation} green full line), which therefore validate an evaluation obtained by an average over previous data measured all by the same-but different than here technique. That is of high importance, given that this validation was missing so far. Furthermore, above $5\,$MeV, where no integral data exist and experimental data are scarce, our results modify \ENDF and \textit{Ev$^{w/o\,this\,work}$} \ensuremath{\overline{\nu}_{p}^{ev}\,} by up to $0.7\%$ and $0.9\%$, respectively. This is consistent with an increasing importance of physics effects leading to biases as \ein increases, which were carefully accounted for in this work. Our data provide therefore a more solid ground for future evaluations.


In conclusion, unrivaled precise and accurate new data on \pu \nubar are reported, which extend the studied range from 1 up to 700 MeV. The data were obtained with the double time-of-flight technique and an innovative setup. It allowed to explicitly   account for experimental systematic bias, which have hampered the precision and accuracy of existing experimental results, thus providing more reliable-than-existing data.
Below 5 MeV a good agreement with the recent \ENDF evaluation is observed validating it, for the very first time, with an independent measurement. A new evaluation performed here with these data shows that they  significantly reduce the uncertainty on evaluated nuclear-data libraries for a nuclide, the \pu, crucial for nuclear energy applications.

With this measurement the experimental challenge of precisely measuring prompt fission-neutron multiplicity on highly radioactive nuclei has been taken up  thanks to the innovative setup and experimental technique. New high-precision \pu \nubar measurements could be realized at incident-neutron energies from 200 keV to 2 MeV, where existing data are highly spread, and even down to 1 keV where no data exist. Moreover these results  open up the possibility of precisely investigating other high-activity actinide nuclei to contribute to a better understanding of the fission process while providing key elements for the development of new technologies relevant for  society.\\

We wish to acknowledge A.~Moens, G.~Sibbens and D.~Vanleeuw from the JRC-Geel target preparation laboratory for providing $^{239}$Pu samples and assisting their mounting in the fission chamber.
We also wish to acknowledge the support of  E.~Bond from LANL C-NR for providing the $^{252}$Cf sample, and of K.~T.~Schmitt from Oak Ridge National Laboratory (formerly of LANL ISR-1).
This work was performed under the auspices of a cooperation agreement between CEA/DAM and DOE/NNSA on fundamental sciences and benefited from the use of the LANSCE accelerator facility.
Work at LANL was carried out under the auspices of the NNSA of the U.S. Department of Energy under contract 89233218CNA000001.
We gratefully acknowledge partial support of the Advanced Simulation and Computing program at LANL and the DOE Nuclear Criticality Safety Program, funded and managed by NNSA  for the DOE.

\bibliography{bibliogr}

\end{document}